\documentstyle[11pt]{article}
\newcommand{\dd}{{\rm d}}
\newcommand{\gsim}{\mathrel{%
   \rlap{\raise 0.511ex \hbox{$>$}}{\lower 0.511ex \hbox{$\sim$}}}}
\newcommand{\lsim}{\mathrel{
   \rlap{\raise 0.511ex \hbox{$<$}}{\lower 0.511ex \hbox{$\sim$}}}}
\def\gam{\gamma}
\def\jj{{\bf j}}
\def\calL{ {\cal L}}  
\def\calT{ {\cal T}} \def\calTi{ \calT_{_\infty} }
\def\calG{ {\cal G}}  
\def\calR{ {\cal R}} 
\def\calV{ {\cal V}}  
\def\calU{ {\cal U}}
\def\calM{ {\cal M}}
\begin{document}

\title{Reflection symmetry breaking scenarios with minimal 
gauge form coupling in brane world cosmology}

\author{Brandon Carter$^1$ and  Jean--Philippe Uzan$^2$}

\date{(1) DARC, CNRS--UMR~8629, Observatoire de Paris, 
F-92195 Meudon (France)
\\(2) Laboratoire de Physique Th\'eorique, CNRS--UMR~8627, B\^at. 210,
\\  Universit\'e Paris XI,    F-91405 Orsay Cedex (France)\\
1 January, 2001}

\maketitle

\begin{abstract}
 This article synthesises and extends recent work on the cosmological
consequences of dropping the usual Z$_2$ reflection symmetry postulate
in brane world scenarios. It is observed that for a cosmological model
of homogeneous isotropic type, the relevant generalised Birkhoff
theorem establishing staticity of the external vacuum in the maximally
symmetric ``bulk'' outside a freely moving world brane will remain
valid for the case of motion that is forced by minimal (generalised
Wess Zumino type) coupling to an external antisymmetric gauge field
provided its kinetic action contribution has the usual homogeneous
quadratic form. This means that the geometry on each side of the brane
worldsheet will still be of the generalised Schwarzschild
anti--de~Sitter type. The usual first integrated Friedmann equation
for the Hubble expansion rate can thereby be straightforwardly
generalised by inclusion of new terms involving 2 extra parameters
respectively measuring the strength of the gauge coupling and the
degree of deviation from reflection symmetry.  Some conceivable
phenomenological implications are briefly outlined, and corresponding
limitations are derived for possible values of relevant parameters.
\end{abstract}

\section{Introduction of minimal coupling}

The proverbially fundamental complementarity of ``brain versus brawn''
has an analogue in the purely mathematical context associated with
names such as Cartan and de Rham wherin gauge equivalence classes of
$p$--forms (i.e.  covariant antisymmetric tensor fields) arise as the
homological complement of $p$--surfaces. Conversely, in the higher
dimensional physical context associated with names such as Polchinski
and Witten, $(p-1)$--branes (i.e.  $p$--surface supported systems)
emerge naturally as concentrated source distributions for gauge
$p$--form fields.  Thus given any $(p-1)$--brane one of the first
questions that comes to mind is what about the corresponding
``brawn'', i.e. the associated gauge $p$--form? In the case of a point
particle (i.e. a zero brane) the answer is usually that it is the
ordinary Maxwellian electromagnetic field.  In the case of the
recently much studied theories that treat our world as a 3--brane in a
5--dimensional ``bulk'', the question of the corresponding gauge
4--form has been generally overlooked, perhaps because the ensuing
``brawn'' force would have to vanish anyway in the reflection
symmetric scenarios that are most commonly~\cite{RS99,BDL00}
considered.

Several authors have discussed models in which the usual Z$_2$ symmetry
postulate in homogeneous isotropic cosmological scenarios treating our
four dimensional world as a 3--brane in a 5--dimensional ``bulk'' was
dropped.  Ida~\cite{ida00} and then Krauss~\cite{krauss00} showed that
such a reflection symmetry breaking can arise due to the choice of
different integration constants for the solution of the vacuum Einstein
equations on either side of the brane, and the general question of the
observable consequences of dropping this symmetry been addressed by
Davis {\it et al.}~\cite{DDPV00}. A different kind of symmetry breaking
has been considered by Deruelle and Dole\v{z}el~\cite{DD00} and
Perkins\cite{perkins00}, who considered bubble walls separating two
distinct five dimensional anti--de~Sitter spacetime domains with
different cosmological constants.  Finally, the possibility that both
kinds of symmetry breaking occur simultaneously has been considered by
Stoica {\em et al.}~\cite{STW00} and by Bowcock {\em et
al.}~\cite{BCG00}, who have shown how to obtain a generalisation of the
usual first integrated Friedmann equation giving the square of the
Hubble expansion rate as a sum of terms in which, as well as the the
ones that are now familiar~\cite{BDEL00} in ``conventional'' (meaning
force free reflection invariant) brane world cosmology there is a new
term (including the extra contribution considered
in~\cite{ida00,krauss00,DDPV00,DD00,perkins00,STW00}) that could have
been important at some stages in the early universe.

The present article provides a synthesis that develops the underlying
physics giving rise to such scenarios by incorporating allowance for
the simplest conceivable coupling to an external antisymmetric
``brawn'' gauge field, $A_{\nu\rho\sigma\tau}$ say (where Greek indices 
run from 0 to 4).  This minimal
coupling case~\cite{CB00} is governed by an action consisting just of
an external kinetic contribution of the usual purely quadratic type,
together with a coupling contribution given by an action density of
the (generalised Wess Zumino) form
\begin{equation}\label{1}  
\calL_{_{\rm co}} =(4!)^{-1} \jj
{^{\,\nu\rho\sigma\tau}} A_{\nu\rho\sigma\tau}\, ,
\end{equation}
for a Dirac distributional source given by some dimensionless coupling
constant $e_{_{ \{4\}}}$ as
\begin{equation}\label{2} 
\jj{^{\,\nu\rho\sigma\tau}} = \overline{\, \jj\,} 
{^{\nu\rho\sigma\tau}}\delta[\zeta] \qquad \hbox{with}\qquad \hskip 1 cm
\overline {\,\jj\,} {^{\nu\rho\sigma\tau}}
=e_{_{ \{4\}}}{\cal E}^{\nu\rho\sigma\tau} \, ,
\end{equation}
where $\zeta$ is a coordinate measuring othogonal distance from the
4--dimensional worldsheet of the 3--brane. Here ${\cal
E}^{\nu\rho\sigma\tau}$ is the antisymmetric tangent element of the
worldsheet normalised as ${\cal E}^{\nu\rho\sigma\tau} {\cal
E}_{\nu\rho\sigma\tau}=-4!$ .  It can be
used, in conjunction with the measure tensor
$\varepsilon_{\mu\nu\rho\sigma\tau}$ of the 5--dimensional background
tensor, to construct the unit normal covector
$\lambda_\mu\equiv\nabla_{\!\mu}\zeta= (4!)^{-1}
\varepsilon_{\mu\nu\rho\sigma\tau}{\cal E}^{\nu\rho\sigma\tau}$.  The
scalar coefficient $e_{_{ \{4\}}}$ must necessarily be uniform over the
worldsheet in order for the source current to satisfy the conservation
law $\nabla_{\!\mu}\jj^{\mu\nu\rho\sigma}=0$ that is necessary for gauge
invariance.  The corresponding ``physical'' (i.e. gauge invariant)
5--form is defined by the exterior derivative of $A_{\nu\rho\sigma\tau}$
as
\begin{equation}\label{3} 
F_{\mu\nu\rho\sigma\tau}\equiv 5 \nabla_{\![\mu}A_{\nu\rho\sigma\tau]}=
F\varepsilon_{\mu\nu\rho\sigma\tau} \, , 
\end{equation}
where $F$ is a pseudo scalar (it is not strictly a scalar because its
sign is parity depenent) in terms of which the purely external 
Lagrangian action governing the gauge field has the form
\begin{equation}\label{4} 
\calL_{_{\rm ex}}= {1\over 2\alpha} F^2 
\end{equation}
where $\alpha$ is a coupling constant. Varying this Lagrangian (\ref{4})
with respect to the 5--dimensional spacetime metric $g_{\mu\nu}$ defines
the external stress energy density tensor ($\calT_{_{\rm
ex}}^{\,\mu\nu}=2\delta \calL_{_{\rm ex}}/\delta g_{\mu\nu}+\calL_{_{\rm
ex}}g^{\mu\nu}$) as
\begin{equation}\label{5}
\calT_{_{\rm ex}}^{\,\mu}{_\nu}=-{1\over 2\alpha} F^2 
g^\mu_{\ \nu}\, .
\end{equation} 
A coupling Lagrangian $\calL_{_{\rm co}}$ of the most general kind
would give rise to another contribution, $\calT_{_{\rm co}}$, to
the stress energy density tensor but this can be shown~\cite{CB00} to 
vanishing in the minimally coupled case (\ref{1}) we are considering here.
article. Varying the coupling and external Lagrangians with respect to
the ``brawn" field $A_{\nu\rho\sigma\tau}$ leads to the field equation
of motion
\begin{equation}\label{6}
\nabla^\mu F_{\mu\nu\rho\sigma\tau}=-\alpha\,
\jj_{\nu\rho\sigma\tau}\, .
\end{equation}
It can thereby be concluded that in the source free region outside the
brane the pseudo scalar field $F$ is uniform, i.e.  that $\nabla_\mu
F=0$.

An important consequence of this is that outside (but not in) the
world brane, the ``brawn'' stress energy tensor (\ref{5}) acts just
as an effective addition to the cosmological constant $\Lambda$ that 
appears in the 5--dimensional Einstein equation according to which
the effect of the total stress energy tensor $\calT_{\mu\nu}$
(including both external and brane contributions) is
given by
\begin{equation}\label{7}  
\calG_{\mu\nu}= \kappa\calT_{\mu\nu}
-\Lambda g_{\mu\nu} \, ,
\end{equation}
in which $\calG_{\mu\nu}\equiv\calR_{\mu\nu}-{1\over 2}\calR
g_{\mu\nu}$, where $\calR_{\mu\nu}$ is the Ricci tensor and $\cal R$ is
its trace. The traditional unrationalised coupling constant
$\kappa$ is related to the quantity $\kappa_{_{(5)}}$ used by several
authors (e.g.~\cite{BDL00}), and to the
5--dimensional analogue $G$ of Newton's constant and its associated
gravitational mass scale $m_{_{\rm G}}$ as defined according to the
appropriate~\cite{ADD98} rationalisation procedure, by
\begin{equation}\label{7a}
\kappa=\kappa_{_{(5)}}^{\,2}=6\pi^2 {\rm G}\, ,\hskip 1 cm
{\rm G}=m_{_{\rm G}}^{-3}\, .
\end{equation}
Thus, in the external region on either side of the brane, substitution
of (\ref{5}) for $\calT_{\mu\nu}$ reduces the Einstein equation
(\ref{7}) to the pseudo vacuum form
\begin{equation}\label{8}  
\calG_{\mu\nu}= -\Lambda^{_\pm} g_{\mu\nu}\, ,
\hskip 1 cm \Lambda^{_\pm}=\Lambda+{\kappa\over 2\alpha}
F^{_\pm 2}\, ,
\end{equation}
where $F^{_+}$ and $F^{_-}$ are the constant values taken by the
pseudoscalar ``brawn'' field in the respective external coordinate
ranges $\zeta>0$ and $\zeta <0$. This minimal coupling case may be
seen as the five dimensional analogous of the standard $p=1$
dimensional case of a charge particle in electromagnetism in which
$e_{_{\{1\}}}$ is the particle charge, and $F_{\mu\nu\rho\sigma\tau}$
is just the analog of the electromagnetic tensor $F_{\mu\nu}$ while
(\ref{6}) is the Maxwell equation.

\section{The generalised (Birkhoff type) bulk staticity theorem}

So long as we are concerned just with cosmological models of
homogeneous isotropic type, the consideration that the external
gravitational field is of pseudo vacuum type allows to invoke the
generalised Birkhoff theorem (see the treatise of Misner {\it et
al.}~\cite{MTW}, whose sign conventions we are following). Its
original version showed that spherical symmetry necessarily entails
the staticity property that was merely assumed as a simplifying ansatz
when Schwarzschild first derived his famous solution. A recent
demonstration with the range of applicability needed in the present
context has recently been provided by Bowcock, Charmousis and
Gregory~\cite{BCG00}, who deal with the case of a 5--dimensional
vacuum metric $\dd s_{\rm V}^2=g_{\mu\nu}\, \dd x^\mu\, \dd x^\nu$
that has the symmetry of a 3--sphere, a 3--plane, or an anti 3--sphere
with metric given, respectively for positive, vanishing, or negative
curvature parameter $k$, by
\begin{equation}\label{9} 
\dd s_{\rm I\!I\!I}^2={\dd\chi^2\over 1-k\chi^2}+\chi^2(\dd\theta^2
+{\rm sin}^2\theta\, \dd\phi^2)\, .
\end{equation}
The conclusion is that wherever equations of the vacuum form 
(\ref{8}) are satisfied, and thus on either side of the (minimally 
coupled) 3--brane under consideration, the metric must be static and 
hence of the generalised Schwarzschild (anti)--de~Sitter form 
\begin{equation}\label{10} 
\dd s_{\rm V}^2= r^2\, \dd s_{\rm I\!I\!I}^2+\frac{\dd r^2}{\calV}-\calV 
\dd t^2\, ,
\end{equation}
with
\begin{equation}\label{11} 
\calV = k -{\Lambda^{_\pm}\over 6} r^2-{2{\rm G}\calM^{_\pm}\over r^2}\, ,
\end{equation}
where $\calM^{_+}$ and $\calM^{_-}$ are constants of integration,
having the dimensions of mass, that characterise the solution in the
respective ranges $\zeta>0$ and $\zeta<0$. (This generalisation is
analogous to that for the charged black hole case, for which the
Birkhoff theorem still applies, giving the static
Reissner--Nordstr\"om solution.) In the compact case $k>0$ the quantity
$r/\sqrt{k}$ is interpretable as a Schwarzschild type radial
coordinate and the mass constants is given by $\calM^{_\pm}= k^2
M^{_\pm}$ where $M^{_+}$ and $M^{_-}$ are interpretable as the masses
of distant source distributions, such as other 3--branes (or in their
absence, of black holes) on either side.  For this positive curvature
case it would commonly be convenient to fix the scale so as to obtain
$k=1$, but such a canonical normalisation is not possible in the
spacially flat case ($k=0$) that is accurate within observational
uncertainties for the large scale representation of our actual
universe.

The fact that the static nature of the external metric is unaffected by
the dynamical evolution of the brane means that there is no way it can
signal its presence to, and thereby perhaps avoids collision with, any
other brane that may be present outside (at a distance whose value as a
function of time is what is commonly referred to as the ``radion'').  The
avoidance of this so called ``radion instability'' problem is one of the
motivations for the development of generalisations~\cite{BM00} involving
the presence in the ``bulk'' of a dilatonic scalar field that can carry
information from one brane to a neighbouring brane. What we have shown
here is that minimal coupling to what is effectively a pseudo scalar
field does not offer any such escape from this ``radion instability'' 
problem. If only a single brane is present the problem of the 
``radion'' as such will not arise, but if either $\calM^{_+}$ or 
$\calM^{_-}$ is non zero (which, as we shall see, is
necessarily the case in the kind of reflection symmetry breaking
scenarios envisaged in~\cite{ida00,krauss00,DDPV00,DD00})  one might
still need to worry about the possibility of colliding with its source
or (if it is source free) of falling into the corresponding black hole.

When one is mainly interested in what is going on in the brane itself
rather than in the ``bulk'' outside, it is commonly convenient to use a
brane based reference system depending on the use of the coordinate
$\zeta$ that measures the orthogonal distance from the worldsheet.  In
the homogeneous isotropic case this gives an expression of the well
known form
\begin{equation}\label{12}
\dd s_{_{\rm V}}^{\, 2}= r^2 \dd s_{_{\rm I\!I\!I}}^{\, 2}+ \dd\zeta^2 -
\nu^2 \dd\tau^2\, ,
\end{equation}
in which the quantities $r$ and $\nu$ are functions just of $\zeta$ and
$\tau$. The latter is normalised in such a way that in the limit
$\zeta\rightarrow 0$ (i.e. on the worldsheet) we have $\nu\rightarrow
1$.  It means that $\tau$ is interpretable as a measure of the proper
cosmological time on the world brane. The corresponding limit
$r\rightarrow a$ defines a function $a(\tau)$ interpretable as a
cosmologically comoving scale factor, whose proper time derivative $\dot
a$ specifies the relevant Hubble expansion rate $\dot a/a$.

Although the two versions (\ref{10}) and (\ref{12}) have long been
separately well known, the relationship between them was not
satisfactorily clarified until the recent work by Mukohyama, Shiromizu,
and Maeda~\cite{MSM00}. These authors showed explicitly how a metric of
the static form (\ref{10}) can be transformed to the brane based version
obtained (in terms of hyperbolic functions of $\zeta$) by Bin\'etruy,
Deffayet, Ellwanger and Langlois~\cite{BDEL00}. The self contained
treatment provided here avoids getting into these explicit
details~\cite{BDEL00,MSM00} by following a more economical approach that
does not depend on the particular (explicitly integrable) form
(\ref{11}) of the function $\calV$, and that could thus be useful in a
broader context (such as might arise if, as well as the ``brawn'' field
$A_{\mu\nu\rho\sigma}$ considered here, a lower order gauge field,
$A_{\mu\nu\rho}$ or $A_{\mu\nu}$ was also present in the bulk).

The essential point is that (in each radial $\{r,t\}$ 2-surface) the distance 
function $\zeta$ is orthogonal to lines of constant $\tau$ that are 
automatically geodesic. The stationarity of  the metric (\ref{10}) therefore 
implies that the unit tangent
vector $x^{\mu\prime}$ (employing the usual convention that dash and dot 
indicate partial differentiation with respect to $\zeta$ and $\tau$ 
respectively) will contract with the time symmetry Killing vector,
as specified by $k_\mu dx^\mu=-d\tau$, to give a constant of the
geodesic motion $E=-k_\mu x^{\mu\prime}$ (that would be intepretable
as energy if instead of being spacelike the geodesic was timelike).
Though independent of $\zeta$, this quantity $E$ depends in general  on
the new time variable $\tau$ whose normalisation it is now convenient to
fix with respect to the rate of change of the scale length $a$ on the
worldsheet by imposing the identification $E=\dot a$. The definition of
$E$ then provides the relation $\dot t=\dot a/\calV$, and hence the unit 
normalisation condition $g_{\mu\nu}x^{\mu\prime} x^{\nu\prime}=1$ takes the form
\begin{equation} \label{13}
\dot a^2= r^{\prime 2}-\calV\, .
\end{equation}
In order to preserve coordinate orthogonality, $\dot x^\mu x^\prime_\mu=0$,
the required transformation, $dt=\dot t \, d\tau+ t^\prime\, d\zeta$ 
and $dr = \dot r\, d\tau+ r^\prime\, d\zeta$, must evidently 
satisfy the condition $\dot r r^\prime=\calV^2 \dot t t^\prime$, 
so substitution in (\ref{10}) now leads directly to {\it q.e.d.}, 
namely the brane based form (\ref{12}), in which the coefficient 
$\nu$ can be seen to be given simply by
\begin{equation} \label{13a}
\nu^2 = \dot  r^2/ \dot a^2\, .
\end{equation}
This short cut derivation of (\ref{12}) has made no use of any specific 
prescription for the dependence of $\calV$ on $r$, but it is now to be
observed that when $\calV$ is given in terms of the relevant generalised
Schwarzschild AdS formula (\ref{11}), the relations (\ref{13}) and 
(\ref{13a}) take precisely the form to which the (bulk) Einstein 
equations were shown to be reducible by Bin\'etruy 
{\it et al.}~\cite{BDEL00}.

The condition of metric continuity on the brane worldsheet (where
$r\rightarrow a$)  is trivially satisfied by 
(\ref{13a}), but (\ref{13}) provides interesting matching conditions.  We
use square and angle brackets respectively for the difference and
average of the values limits on opposite sides, so that in particular
for the effective cosmological constant we have
\begin{equation}\label{14} 
[\Lambda]\equiv\Lambda^{_+}-\Lambda^{-}=
{\kappa\over \alpha} \langle F\rangle[F\big]
\, ,\hskip 1 cm
\langle \Lambda\rangle\equiv{_1\over^2}\Big(\Lambda^{+}+\Lambda^{-}\Big)
=\Lambda +{\kappa\over 2 \alpha}\langle F^2\rangle\, .
\end{equation}
One finds that subtraction of (\ref{13}) gives the  relation 
\begin{equation} 
\langle r^\prime\rangle [r^\prime]
=-a^2 {[\Lambda]\over 12}-{{\rm G}[\calM]\over a^2} \, ,\label{15}
\end{equation}
which is of no interest in the Z$_2$ reflection symmetric case where all
the terms vanish, while averaging (\ref{13}) gives the more
generally useful expansion rate formula
\begin{equation}\label{16}
\left({\dot a\over a}\right)^2= {\langle\Lambda\rangle\over 6}+
{\langle r^{\prime 2}\rangle -k \over a^2} +
{2{\rm G}\langle\calM\rangle\over a^4}\, .
\end{equation}

\section{Solution of the Junction conditions}

To determine the evolution of the system we are interested in, we 
need to use the well known~\cite{MTW} Darmois Israel junction
conditions according to which the active gravitational effect
of the surface stress energy density $\overline \calT^{\mu\nu}$
of the brane is governed by the equation 
\begin{equation}\label{17}
\big[K_{\mu\nu}\big] -\big[ K \big]\gam_{\mu\nu}=
\kappa \overline{\cal T}_{\mu\nu}\, 
\end{equation}
where $\gam_{\mu\nu}$ and $K_{\mu\nu}$ are respectively the first and
the second fundamental forms ($K$ being its the trace), defined as
\begin{equation}\label{22}
\gam_{\mu\nu}\equiv g_{\mu\nu}-\lambda_\mu\lambda_\nu\, 
,\hskip 1 cm
K_{\mu\nu}\equiv -\gamma^\rho_{\, \nu}\nabla_{\!\mu}\lambda_\rho \, ,
\end{equation}
$\lambda_\mu=\nabla_{\!\mu}\zeta$ being the unit
(i.e. $\lambda_\rho\lambda^\rho=1$) vector
normal to the brane (i.e $\gam_{\mu\nu}\lambda^\mu=0$).

It is also necessary to solve the equations governing the passive 
evolution for the worldsheet, which has often been worked out {\em ad hoc} 
in particular applications, but which can also be expressed~\cite{CB00} 
in a simple generally valid form as
\begin{equation}\label{18} 
\overline \calT{^{\mu\nu}}\langle K_{\mu\nu}\rangle =\overline f\, ,
\end{equation}
where $\overline f$ is the force density acting on the brane. This will
be given in terms of the contraction of the external (bulk) stress 
energy density as
\begin{equation} \label{19}
\overline f =-\lambda_\mu\lambda_\nu\big[\calT_{_{\rm ex}}^{\!\mu\nu}
\big]\, .
\end{equation} 
This must of course vanish in the reflection symmetric configurations
that are most commonly considered, but need not do so in the more
general context considered here. It can be seen from (\ref{6}) and
from the expression (\ref{5}) for $F_{\mu\nu\rho\sigma\tau}$ that
$\lambda_\mu\nabla^\mu F=\alpha e_{_{\{4\}}}\delta[\zeta]$, which
implies that
\begin{equation}\label{20} 
[F]=\alpha e_{_{\{4\}}}\, .
\end{equation}
The force density is found to have a constant but generically non 
vanishing value given by
\begin{equation}
\overline f =e{_{_{\{4\}}}}\langle F\rangle =
\kappa^{-1} [\Lambda] \, .\label{21}
\end{equation} 

In order to apply the preceeding formulae one has to work out the exact
expressions of the first and second fundamental forms as defined in
equation~(\ref{22}). The first fundamental form is obviously 
given by
\begin{equation}\label{23}
\dd s_{_{\rm IV}}^2=\gam_{\mu\nu}\,\dd x^\mu\,\dd x^\nu=a^2
\dd s_{\rm I\!I\!I}^2-\dd\tau^2\, ,
\end{equation}
and the second fundamental form by
\begin{equation}\label{24} 
K_{\mu\nu}\,\dd x^\mu\,\dd x^\nu=-a r^\prime 
\dd s_{\rm I\!I\!I}^2+\nu^\prime \dd\tau^2\, .
\end{equation}

We also need to know the form of the worldsheet stress energy density
tensor $\overline\calT{^{\mu\nu}}$. To be compatible with the hypothesis
of a homogeneous isotropic geometry, $\overline\calT{^{\mu\nu}}$ must itself
be of the homogeneous isotropic form, i.e.
\begin{equation}\label{25}   
\overline\calT{^{\mu\nu}} =\calU u^\mu u^\nu-\calT\left(
\gam^{\mu\nu} + u^\mu u^\nu\right)
\end{equation}
with respect to the preferred unit vector defined by
$u_\mu=-\nabla_{\!\mu}\tau$. Here $\calU$ is the total energy density and
$\calT$ is the total brane tension. It is normally assumed that the
directly observable energy density $\rho$ and pressure $P$ (i.e. of the
cosmic fluid) represent small deviations from an isotropic
(inflationary) limit state given by $\overline\calT{^{\mu\nu}}=-\calTi
\gam^{\mu\nu}$ where $\calTi$ is a fixed tension 
(approached by both $\cal T$ and $\calU$ in the limit 
$a\rightarrow\infty$ ) in terms of which the actual tension and 
energy density are given by
\begin{equation}\label{26} 
\calT\equiv\calTi-P\, ,\hskip 1 cm \calU\equiv\calTi+\rho\, .
\end{equation}
Substitution of (\ref{25}) in (\ref{17}) gives explicit jump conditions
of the well known~\cite{BDL00} form
\begin{equation}\label{27} 
[r^\prime]=- {a\kappa\over 3}\calU\, ,\hskip 1 cm
[\nu^\prime]={\kappa\over 3}\left(2\calU-3\calT\right)\, .
\end{equation}
Now, from (\ref{18}) we obtain the relation
\begin{equation}\label{28} 
\calU \langle\nu^\prime\rangle +3\calT {\langle r^\prime\rangle \over
a}=\overline f \, ,
\end{equation}
which differs from the usual version by the presence of the force
density term on the right.

Using the expression (\ref{27}) for $[r']$, we obtain, from (\ref{15}),
that $2\langle r^\prime\rangle$ (which is the same as the quantity ``$d$''
introduced as a measure of the deviation from reflection symmetry by
Davis {\it et al.}~\cite{DDPV00}) is given by
\begin{equation}\label{29} 2\langle r^\prime\rangle=
{3\over\kappa\calU} \left({a[\Lambda]\over 6}+ {2{\rm
G}[\calM]\over a^3}\right)\, .  
\end{equation} 
This generalises the corresponding formula of Davis {\it et al.}  by the 
inclusion of the $[\Lambda]$ term representing the effect of the ``brawn'' 
force, and it also provides a more specific interpretation of their ``$F$'' 
term by showing that (as observed by Ida~\cite{ida00}) it is proportional 
to the (reflection symmetry breaking) difference $[\calM]$ between 
the mass terms on opposite sides, and hence that it would have to be absent
if both mass terms were zero (as was assumed in the work of  Deruelle 
and Dole\v{z}el~\cite{DD00} and Perkins~\cite{perkins00}). It is
to be observed that the gravitational coupling coefficients can be
 cancelled out of (\ref{29}) and hence also out of the expressions
for the components of the mean value of the first fundamental form 
(\ref{24}), which vanish in the reflection symmetric case, and
which will be given in the general case by 
\begin{equation}\label{29b}
\calU\langle r^\prime\rangle= {[\calM]\over 2\pi^2 a^3}+
{a\over 4} e_{_{\{4\}}}\langle F\rangle\, ,\hskip 1 cm
\calU\langle \nu^\prime\rangle= -{3[\calM]\calT\over 2\pi^2 a^4\calU}+
\Big(1 -{3\calT\over 4\calU}\Big) e_{_{\{4\}}}\langle F\rangle\, .
\end{equation}

Using (\ref{29}) and the first equation of (\ref{27}) respectively for
$\langle r^\prime\rangle$ and $[r^\prime]$ and using the decomposition
$\langle r^{\prime 2}\rangle$ $=\langle
r^\prime\rangle^2+[r^\prime]^2/4$ to evaluate $\langle r^{\prime
2}\rangle$ in (\ref{16}), we can now proceed directly to the relevant
generalisation of the Friedmann equation, which takes the form
\begin{equation}\label{30} \left({\dot a\over
a}\right)^2=\left({\kappa\calU\over 6} \right)^2
+{\langle\Lambda\rangle\over 6} -{k\over a^2} +{2{\rm G}\langle\calM
\rangle\over a^4} +{1\over\left(2\kappa\calU\right)^2}
\left({[\Lambda]\over 2}+{6{\rm G}[\calM]\over a^4}\right)^2 \, .
\end{equation} 
This agrees with what was obtained in a rather more circuitous manner by
Stoica {\it et al}~\cite{STW00}, and by Bowcock {\it et al.}~\cite{BCG00}, 
who were the first to undertake a systematic investigation of reflection 
symmetry breaking scenarios with the degree of generality considered 
here, but who used a system in which the 5--dimensional gravitational 
coupling constant $\kappa$ was set to unity, thereby obscuring the way it 
affects the various terms. As well as directness, the present approach has 
the advantage of showing more explicitly how the various coupling 
coefficients contribute, and in particular how $\kappa$ actually cancels 
out of the formulae (\ref{29b}) and, as will be demonstrated immediately 
below, from last term of (\ref{30}) which is the only one whose presence 
depends on reflection symmetry breaking.

\section{Investigation of the parameter space}

In order to interpret the Friedmann--like equation (\ref{30}), we use
(\ref{26}) to rewrite it in terms of the ratio $\varepsilon\equiv
\rho/\calTi$, as
\begin{eqnarray}\label{40}
\left({\dot a\over a}\right)^2&=&{8\pi{\rm G}_{_{4}}\over 3}\rho 
-{k\over a^2}+\frac{\Lambda_{_4}}{3}+{2{\rm G}\langle\calM\rangle\over a^4}
+\big(\pi^2{\rm G}\rho\big)^2  \nonumber\\
&+& {(1+\varepsilon)^{-2}\over(2\calTi)^2}
\left\lbrace {[\calM]\over\pi^2 a^4}\left(
{[\calM]\over\pi^2a^4}+e_{_{\{4\}}}\langle F\rangle\right)
+\Big({ e_{_{\{4\}}}\langle F\rangle\over 2}\Big)^2
\big(3+2\varepsilon\big)\varepsilon^2\right\rbrace 
\end{eqnarray}
in which the first three terms of the r.h.s. are those of the standard
Friedmann equation, with the effective 4--dimensional Newton constant
${\rm G}_{_4}$ and the effective 4--dimensional cosmological constant
$\Lambda_{_4}$ given by
\begin{equation}\label{32}
8\pi{\rm G}_{_{4}}={6\over \calTi}\left( \left(\pi^2{\rm G}\calTi\right)^2
 -\left({e_{_{\{4\}}}\langle F\rangle\over 4\calTi}\right)^2\right) \, ,
\end{equation}
and
\begin{equation}\label{33} 
2\Lambda_{_4}=\langle\Lambda\rangle+ 6
\left( \left(\pi^2{\rm G}\calTi\right)^2
 +\Big({e_{_{\{4\}}}\langle F\rangle\over 
4\calTi}\Big)^2\right)\, . 
\end{equation}

In terms of the brane mass scale given by $m_{_\infty}^{\,4}=\calTi$, and
of the ordinary Planck mass scale  given by $m_{_{\rm P}}^{-2}\equiv{\rm
G}_{_4}$, it is convenient to parametrise the relative importance in
(\ref{32}--\ref{33}) of the gauge field contribution $\langle F\rangle$
and of the bulk gravitational coupling constant given by $m_{_{\rm
G}}^{\,-3}\equiv{\rm G}$, in terms of a dimensionless hyperbolic angle
$\chi$ such that
\begin{equation}\label{32b} 
e_{_{\{4\}}}\langle F\rangle=\left({\pi\over 3}
\right)^{1/2}\ {8 m_{_\infty}^{\ 6}\, {\rm sh}\,\chi\over m_{_{\rm P}}}\, ,
\hskip 1 cm {\rm G}=\left({\pi\over 3}\right)^{1/2}{2 \,{\rm ch}\,\chi
\over \pi^2 m_{_\infty}^{\,2} m_{_{\rm P}}}\, .
\end{equation}
Besides the three independent parameters involved in (\ref{32b}),
namely $\chi$, $m_{_\infty}$ and $m_{_{\rm P}}$, of which only the
latter is experimentally known in advance, this model involves four
more independent parameters which are the bulk mass scales
$\calM^{_\pm}$, the ordinary cosmological constant $\Lambda_{_{4}}$
and the cosmological curvature scale $k$. Only the two latter are
roughly known in advance, neither of them differing from zero by an
amount that can be reliably measured yet. Since $\Lambda_{_{4}}$ is at
most of the order of the present day cosmological closure value (i.e.
$m_{_{\Lambda_{4}}}\lsim10^{-60}m_{_{\rm P}}$), it is effectively
negligible for our present purpose, i.e. we can set
$\Lambda_{_{4}}=0$. This means that the effective bulk cosmological
constants and the corresponding curvature length and mass scales
$\ell_{_\Lambda}= m^{-1}_{_\Lambda}$ as defined by
$-\Lambda\equiv6\ell^{-2}_{_\Lambda}$ on either side of the brane 
will be given (using  (\ref{32b}) for  $\langle\Lambda\rangle$,  and 
using (\ref{21}) and (\ref{32b}) for  $[\Lambda]$ ) by
\begin{equation}\label{33b} 
-\Lambda^{_\pm}=
6m^{_\pm\, 2}_{_\Lambda}= 8\pi\frac{m_{_\infty}^4}{m_{_{\rm P}}^2}
\,{\rm e}^{\mp 2\chi}\, , \hskip 1 cm \ell^{_\pm}_{_\Lambda}=
\left({3\over\pi}\right)^{1/2}{m_{_{\rm P}}\over 2m_{_\infty}^{\,2}}
\,{\rm e}^{\pm\chi}\, .\end{equation}

In order to place some limits on the four independent unknown
constants $\calT_{_\infty}$, $\chi$, and the bulk mass constants it is
convenient to replace $\calM^{_\pm}$ by variable mass parameters
\begin{equation}\label{40m} 
\mu^{_\pm}\equiv 2{\rm G}\calM^{_\pm}
\left( 1 \pm { {\rm th}\, \chi\over(1+\varepsilon)^2}\right)\, .
\end{equation}
With this notation, we can convert (\ref{40}) to the form
\begin{equation}\label{40b}
\left({\dot a\over a}\right)^2={8\pi\rho\over 3 m_{_{\rm P}}^2}
-\frac{k}{a^2}+{\Lambda_{_4}\over 3} +\Delta\, ,
\end{equation}
in which the final term $\Delta$ represents the deviation from the
traditional Friedmann equation and is expressible as
\begin{equation}\label{40c}
\Delta= \left(\pi^2{\rm G}\rho\right)^2\left(1 +{(3+2\varepsilon)\,
{\rm th}^2\chi\over (1+\varepsilon)^2}\right)
+\left( {[\calM]\over 2\pi^2 a^4\calT_{_\infty}(1+\varepsilon)}\right)^2
+ {\langle\mu\rangle\over a^4}
\, ,
\end{equation}
in which the first two terms are manifestly positive definite.
Furthemore if (unlike some authors who, in scenarios involving more
than one brane, have envisaged negative mass densities) we make the
traditional assumption that all external mass distributions must be
positive (hence $\calM^{_\pm}\geq0$), then it can be deduced from the
expression (\ref{40m}) that both  variables $\mu^{_+}$ and $\mu^{_-}$ must
also be everywhere positive, though in the case of the latter only
marginally if $\chi$ is large.

Before proceeding to exploit these positivity properties for the
purpose of placing limits on the parameters involved, we need to start
by choosing a convenient normalisation for the scale factor $a$. Since
the spacial curvature factor $k$ is too small to have been reliably
measured yet ($k=0$ being a plausible possibility and certainly a good
approximation), the mathematically convenient choice $k=1$ is in
practice unavailable.  Whatever $k$ may be, a physically convenient
choice is to normalise the present value of the scale factor $a$ to
agree with the length scale characterising the cosmological background
radiation temperature $\Theta$, i.e. to simply to take
$a=\Theta^{-1}$.  This normalisation has the advantage that it holds
when extrapolated backwards in time so long as photon creation
processes remain insignificant, which means ever since the temperature
dropped below the electon positron pair creation threshold given by
$\Theta\approx 2m_{\rm e}$.
 
In particular the temperature $\Theta_{_{\rm N}}$ of nucleosynthesis,
which is of greatest interest for the purpose of obtaining rigourous
observational constraints, lies in the range $0<\Theta<
2m_{_{\rm e}}$ in which the product $a\Theta$ remains constant.
Moreover throughout the radiation dominated era during which
\begin{equation}\label{34}
\rho =\frac{\pi^2 g_*}{30} \, \Theta^4
\end{equation}
the order of magnitude relation $\Theta\approx a^{-1}$ will still be
valid as a useful approximation so long as the dimensionless coefficient
$g_*$ representing the effective number of relativistic degrees of
freedom remains comparable with its usual value $g_*\simeq 10.75$ at
nucleosynthesis.  Under such conditions, with $k$ and $\Lambda_{_{\rm
4}}$ now set to zero (since they are negligible in that era), one can
rewrite (\ref{40b}) as
\begin{equation}\label{41c}
\left({\dot\Theta\over\Theta}\right)^2={8\pi\rho\over 3 m_{_{\rm
P}}^2} +\Delta\, .
\end{equation} 
The deviation term (\ref{40c}) can be
expressed in terms of correspondingly normalised ``bulk'' mass scales
$m_{_{\rm B}}^{\,_\pm}\equiv\calM^{_\pm}/2\pi^2$ (representing the
distant external mass associated with a small comoving volume of size
$a^3)$ as
\begin{equation}\label{41d}
\Delta={4\pi\over 3\calT_{_\infty}} \left({\rho\,{\rm ch}\,\chi\over
m_{_{\rm P}}}\right)^2 \left(1 +{(3+2\varepsilon)\, {\rm
th}^2\chi\over (1+\varepsilon)^2}\right) + \Big({[m_{_{\rm
B}}]\Theta^4\over\calT_{_\infty}(1+\varepsilon)}\Big)^2
+{\mu^{_+}\Theta^4\over 2}+ {\mu^{_-}\Theta^4\over 2}\, .
\end{equation}
Since each terms is positive, the condition that $\Delta$ should 
be small compared with the leading term in (\ref{41c}) when
$\Theta\approx\Theta_{_{\rm N}}$ imposes this smallness
requirement on each of the four separate terms in (\ref{41d}). The 
constraint on the first term gives the basic requirement that
\begin{equation}\label{50} 
\Theta_{_{\rm N}}^{\,2}\, {\rm ch}\, \chi\, \ll 
\sqrt{\frac{15}{2\pi^2 g_*}}\,m_{_\infty}^{\, 2}\, ,
\end{equation}
with the convenient corollary that the value of $\varepsilon$ at
the time of nucleosynthesis satisfies $\varepsilon_{_{\rm N}}\ll 1$
throughout the range $\Theta\leq \Theta_{_{\rm N}}$ (and considerably
beyond if $\chi$ is large). From the second term, one
obtains the requirement
\begin{equation}\label{51}
\left\vert[m_{_{\rm B}}]\right\vert\ll \sqrt{\frac{4\pi^3 g_*}{45}}\,
{m_{_\infty}^{\ 4}\over m_{_{\rm P}}\Theta_{_{\rm N}}^{\,2}}\, .
\end{equation}
One can understand these two constraints by expanding $\Delta$ in
series with respect to $\varepsilon\ll1$; they simply state that the
terms scaling faster than $\Theta^4$ (e.g. $\Theta^8$\ldots) are
negligible at the time of nucleosynthesis, thus explaining the
dependence on $\Theta_{_{\rm N}}$ in (\ref{50}) and (\ref{51}).
The third and fourth terms in (\ref{41d}) scale as radiation, thus
behaving like extra relativistic degrees of freedom which are
constrained by the fact that $g_*$ cannot deviate for more than 20\%
from its expected value $g_*=10.75$. If we choose an
orientation convention so that $e_{_{\{4\}}}\langle F\rangle$ is
positive (thus implying that $\chi\geq 0$) then the constraint on the
third term of (\ref{41d}) gives the requirement
\begin{equation}\label{52} 
m_{_{\rm B}}^{_+}\, {\rm ch}\, \chi \ll
\frac{2\pi^2 g_*}{15}\sqrt{\frac{\pi}{3}}
{m_{_\infty}^{\, 2}\over m_{_{\rm P}}}\, .
\end{equation}
Due to the possibility of partial cancellation, the condition 
obtained from the last term in (\ref{41d}) is less restrictive:
by (\ref{50}), since  $\varepsilon_{_{\rm N}}{\rm ch}^2\chi\ll1$, 
it is expressible as
\begin{equation}\label{53}
{m_{_{\rm B}}^{_-}\over {\rm ch}\,\chi}\ll
\frac{2\pi^2 g_*}{15}\sqrt{\frac{\pi}{3}}
{m_{_\infty}^{\, 2}\over m_{_{\rm P}}}\, .
\end{equation}
As expected, these two conditions are independent of $\Theta_{_{\rm
N}}$ since the contributions of the terms in $\mu^{_\pm}$ are 
proportional to that of the radiation.

Using that $\left\vert[m_{_{\rm B}}]\right\vert<m_{_{\rm B}}^{_+}+
m_{_{\rm B}}^{_-}$ and that ${\rm ch}\,\chi+1/{\rm ch}\,\chi<2{\rm ch}\,\chi$,
we deduce that $\left\vert[m_{_{\rm B}}]\right\vert<4\pi^2
\sqrt{\pi/3}{m_{_\infty}^{\, 2}/(15m_{_{\rm P}}})$, and thus that the
condition (\ref{50}) on $\chi$ leads to the constraint (\ref{51}).
This means that (\ref{51}) is not needed as an independent
condition, but is an automatic consequence of the other three conditions
(\ref{50}), (\ref{52}) and (\ref{53}).

Furthermore experimental limits on deviations from the Newtonian
law of gravity imply an upper bound~\cite{limit2} on the length scale
$\ell_{_\Lambda}$, and a corresponding lower
bound on the mass scale $m_{_\Lambda}$ in the bulk. Since no such 
deviation has been detected above a millimeter~\cite{limit}, the present 
bound is $m_{_\Lambda}\gsim10^{-3}$eV, which happens to be comparable 
with present temperature $\Theta_{_0}$ of the cosmic 
microwave background, and can thus be written as
\begin{equation}\label{54}
m_{_\Lambda}\gsim\Theta_{_0}\, .
\end{equation}
Previous discussions~\cite{limit2} of this bound considered only 
scenarios characterised by Z$_2$ reflection symmetry. When this symmetry 
is broken, the bound on the cosmological length scale, and by (\ref{27} 
and (\ref{29b}) on the corresponding asymptotic curvature limit 
$K^{_\pm}_{\,\mu\nu} \sim \pm m^{_\pm}_{_\Lambda}\gamma_{\mu\nu}$
as $a\rightarrow\infty$, must presumably be satisfied on each
side separately. Since our orientation convention is such that
$\ell^{_+}_{_\Lambda}$ is greater than $\ell^{_-}_{_\Lambda}$ we
deduce that in the generic case the mass values given by (\ref{33b}) 
will be characterised by
\begin{equation}\label{54b}
m^{_-}_{_\Lambda} \geq m^{_+}_{_\Lambda}\gsim\Theta_{_0}\, .
\end{equation}

\section{Implications and conclusions}

The requirement (\ref{54b}) is evidently most severe for what, 
according to our convention, is the negative side of the brane,
so by (\ref{33b}) it gives an order of magnitude restriction
expressible as
\begin{equation}\label{55}
 m_{_\infty}^{\, 2}\gsim \sqrt{{3\over\pi}} \Theta_{_0} m_{_{\rm P}}
\,{\rm ch}\,\chi\, .
\end{equation}
This is evidently strong enough to ensure that (\ref{50}) will be
satisfied as an automatic consequence.
Nucleosynthesis and laboratory experiments on the gravitational force
thus provide just three independent constraints. One of them is 
given by (\ref{55}), and the two others are (\ref{52}) and 
(\ref{53}), which can be combined as
\begin{equation}\label{55a}
m_{_{\rm B}}^{_\pm}\ll \frac{2\pi^2}{15}\sqrt{\frac{\pi}{3}}
{m_{_\infty}^{\, 2}\over m_{_{\rm P}}}\left({\rm ch}
\,\chi\right)^{\mp1}\, .
\end{equation}
If $m_{_\infty}$ has the minimum value allowed by
(\ref{55}), i.e.  $m_{_\infty}^{\, 2}$ $\approx
(3/\pi)^{1/2}\Theta_{_0} m_{_{\rm P}} \,{\rm ch}\,\chi$, then
(\ref{55a}) simply gives
\begin{equation}\label{55b}
m_{_{\rm B}}^{_+}\ll{2\pi^2\over 15}\Theta_{_0}\, ,\hskip 1 cm
m_{_{\rm B}}^{_-}\ll{2\pi^2\over 15}\Theta_{_0}({\rm ch}\,\chi)^2\, .
\end{equation}

To estimate the orders of magnitude imposed by the three constraints
(\ref{55}) and (\ref{55a}),  we use the approximate values
$\Theta_{_0}\sim2\times10^{-13}\,$GeV, $m_{_{\rm P}}\sim10^{19}\,$GeV
and $\Theta_{_{\rm N}}\sim1\,$MeV, respectively for the cosmic
background radiation temperature, the Planck mass and the
nucleosynthesis temperature. 
It is convenient to distinguish two regimes as follows.

\begin{enumerate}
\item Type I scenarios, characterised by $\chi\lsim1$: in these scenarios 
the contribution $e_{_{\{4\}}}\langle F\rangle$ is relatively unimportant
both in (\ref{32}) and (\ref{33}). Using ${\rm
ch}\,\chi\simeq1$, we deduce from (\ref{55}) that the mass scale
specifying the limiting value
of the brane tension, $\calT_{_\infty}=m_{_\infty}^{\,4}$, must satisfy
\begin{equation}\label{70}
m_{_\infty}\gsim 10^6
\Theta_{_{\rm N}}\approx1\,\hbox{TeV}\, ,
\end{equation}
which is the same as in the usual Z$_2$ reflection symmetric scenarios. 
In this case, from (\ref{55a}), the two external mass scales satisfy 
the same constraint
\begin{equation} \label{70a}
m_{_{\rm B}}^{_\pm}\ll
\frac{2\pi^2}{15}\sqrt{\frac{\pi}{3}} {m_{_\infty}^{\, 2}\over
m_{_{\rm P}}}\, ,
\end{equation}\label{59}
which in the limit $m_{_\infty}\approx 1\, \hbox{TeV}$  gives
$m_{_{\rm B}}^{_\pm}\ll 10^{-10}\,\hbox{MeV}$.

\item Type II scenarios, characterised by $\chi\gg1$: in these
scenarios the contribution $e_{_{\{4\}}}\langle F\rangle$ from the
gauge field is very large, so there has to be a a fine tuning between
its value and that of the correspondingly large five dimensional
gravitational constant ${\rm G}$ to as to get the relatively small
observed value of ${\rm G}_{_4}$ by (\ref{32}).  In this case
(\ref{70}) is to be replaced by a more restrictive condition of the
form
\begin{equation}\label{71}
 m_{_\infty} \gsim ({\rm ch}\,\chi)^{1/2}\, \hbox{TeV} \,
.\end{equation} 
In such scenarios the external mass scales satisfy
different constraints. In particular if the brane mass scale has the
minimum value $m_{_\infty} \approx ({\rm ch}\,\chi)^{1/2}\,
\hbox{TeV}$ allowed by (\ref{71}) we get $m_{_{\rm B}}^{_+}\ll
10^{-10}\,\hbox{MeV}$ which is the same as before, but on the other
side (\ref{55a}) gives the potentially much less restrictive condition
$m_{_{\rm B}}^{_-}\ll 10^{-10} ({\rm ch}\, \chi)^2\,\hbox{MeV}$.
\end{enumerate}

To sum up, in this article we have introduced the simplest conceivable
coupling to an external gauge field, which has been shown to be
compatible with staticity. Like the effect~\cite{ida00,krauss00,DDPV00}
of external masses, this ``brawn'' coupling breaks the Z$_2$ reflection
symmetry and gives a mechanism for the set up considered for instance
by Deruelle and Dole\v{z}el~\cite{DD00}, Stoica {\em et
al.}~\cite{STW00}, Perkins~\cite{perkins00}, Davis {\em et
al.}~\cite{DDPV00} and Bowcock {\em et al.}~\cite{BCG00}. We have
given a simple self contained derivation of the generalised Friedmann
equation which now depends on five adjustable parameters, and we have used 
its Friedmannian limit to identify the four dimensional Newton constant
and cosmological constant. The 4 independent
parameters involved, namely the hyperbolic angle $\chi$, 
the brane tension mass scale $m_{_\infty}$, and the external masses per 
thermal volume $m_{_{\rm B}}^{_+}$ and $m_{_{\rm B}}^{_-}$
(2 more than are needed for the reflection symmetric
case in which $\chi=0$ and  $m_{_{\rm B}}^{_+}=m_{_{\rm B}}^{_-}$) have
been shown to be subjet to 3 independent constraints (one  more than in the 
symmetric case) of which two arise from  nucleosynthesis measurements and 
one from gravitational laboratory experiments above a millimeter.  
We distinguish two kinds of
regime  depending on the value of the hyperbolic angle $\chi$ that
characterises the relative importance of the ``brawn'' field. In 
the type II case for which this parameter is large  
 the external mass scale can (on one but not both sides) 
be much larger than in the usual reflection symmetric case,
but for this to be possible the brane mass scale $m_{_\infty}$ must also
be much larger that its usual lower limit of the order of a TeV.


\end{document}